\documentclass[12pt]{article}

\usepackage{graphicx}
\usepackage{amsmath}
\usepackage{amssymb}
\usepackage{dsfont}
\usepackage{hyperref}

\usepackage{caption}
\usepackage{titlesec}
\titleformat{\section}
  {\normalfont\fontsize{15}{18}\bfseries}{\thesection}{1em}{}
\newcommand{\poiss}[1]{\left\{#1\right\}}
\newcommand{\lp}[0]{\nonumber \\}

\begin{document}

${}$\\
\begin{center}
\vspace{36pt}
{\Large \bf More on ``Little Lambda'' in \\ \vspace{12pt} 
Ho\v{r}ava-Lifshitz Gravity}
\vspace{48pt}

{\sl R. Loll$\,^{a,b}$}
and {\sl L. Pires$\,^{a}$}

\vspace{24pt}

{\footnotesize

$^a$~Radboud University Nijmegen, \\
Institute for Mathematics, Astrophysics and Particle Physics, \\
Heyendaalseweg 135, NL-6525 AJ Nijmegen, The Netherlands.\\ \vspace{2pt} 
{email:  r.loll@science.ru.nl, l.pires@science.ru.nl}\\

\vspace{10pt}

$^b$~Perimeter Institute for Theoretical Physics,\\
31 Caroline St N, Waterloo, Ontario N2L 2Y5, Canada.\\
{email: rloll@perimeterinstitute.ca}
}

\vspace{48pt}

\end{center}

\begin{center}
{\bf Abstract}
\end{center}
We analyze different claims on the role of the coupling constant $\lambda$ in so-called $\lambda$-$R$ 
models, a minimal generalization of general relativity inspired by Ho\v{r}ava-Lifshitz gravity. 
The dimensionless parameter $\lambda$ appears in the kinetic term of the Einstein-Hilbert action, leading
to a one-parameter family of classical theories.
Performing a canonical constraint analysis for closed spatial hypersurfaces, 
we obtain a result analogous to that of Bellor\'in and Restuccia, who showed
that all non-projectable $\lambda$-$R$ models are {\it equivalent} to general relativity in the asymptotically flat case.
However, the tertiary constraint present for closed boundary conditions assumes a more general form. 
We juxtapose this with an earlier
finding by Giulini and Kiefer, who ruled out a range of $\lambda$-$R$ models by a physical, cosmological 
argument. We show that
their analysis can be interpreted consistently within the {\it projectable} sector of Ho\v{r}ava-Lifshitz gravity, thus
resolving the apparent contradiction. 

\newpage

\section{Introduction}
In absence of a reliable phenomenology for probing physics at the Planck scale,
one of the few tests we have for nonperturbative candidate theories of quantum gravity is whether or not they
can reproduce (aspects of) general relativity in a suitable low-energy limit. 
Depending on how the quantum theory is given, one needs to apply some care when comparing it 
to a continuum formulation of classical gravity, because the latter typically carries a redundancy as a result of its
invariance under spacetime diffeomorphisms. For example, if the quantum theory effectively contains a (partial)
gauge fixing, it may not be appropriate to compare the functional form of its action with the standard action of general 
relativity in terms of metric variables, say. Of course, to avoid such complications any comparison should be phrased in terms
of observables, but in gravity these are often difficult to come by.   

This consideration is relevant in Ho\v{r}ava-Lifshitz gravity \cite{hor}, whose action in terms of metric variables differs from
the Einstein-Hilbert action even in the infrared limit, that is, considering only terms at most quadratic in spatial derivatives. 
The difference occurs because the action of Ho\v{r}ava-Lifshitz gravity is not invariant under four-dimensional diffeomorphisms. 
This reflects the theory's key assumption of the existence of an ultraviolet fixed point at which time and space scale differently, 
leading to a reduced symmetry group that can accommodate this property. 
The simplest choice of such a group, implemented in \cite{hor}, is given by the foliation-preserving diffeomorphisms, 
consisting of three-dimensional spatial diffeomorphisms together with time reparametrizations.\footnote{In principle, 
other choices for the initial field content and symmetry group are possible, see \cite{hor2} for an example.} 
Under this symmetry, the kinetic term of the action acquires an extra coupling -- the ``little lambda" of our title -- since 
both the square $K_{ij}K^{ij}$ 
of the extrinsic curvature tensor and its trace-squared $K^2$ are separately invariant under the reduced symmetry group, 
whereas in general relativity only the precise 
linear combination $K_{ij}K^{ij}\! -\! K^2$ is invariant under four-diffeomorphisms.

Despite having a different symmetry group, Ho\v{r}ava-Lifshitz gravity (HLG) is formulated in terms of the usual 
metric degrees of freedom of gravity, written in the ADM formulation \cite{adm}. 
Comparing its low-energy limit with that of general relativity (GR) is therefore relatively straightforward, if one
keeps in mind our earlier remark that a different functional form of the action does not necessarily signal an
inequivalent theory. There are two apparently contradictory results \cite{kief,ven} on the equivalence or otherwise of the low-energy limits 
of HLG and GR, which the present work aims to resolve. As we will see, they are due to some subtleties in the comparison, 
including the fact that there are several versions of Ho\v{r}ava-Lifshitz gravity, whose classical limits are not the same. 

Our presentation will proceed as follows. To make the treatment self-contained, we recap in Sec.\ \ref{hl_recap} the ingredients 
of Ho\v{r}ava-Lifshitz gravity that will be necessary in our classical investigation of the $\lambda$-$R$ models. Sec.\ \ref{np_theory} 
contains our analysis of the constrained structure of the $\lambda$-$R$ model derived from non-projectable HLG, for closed spatial
slices. This involves a careful re-examination of all steps made in the asymptotically flat case, which in \cite{ven} was shown to be 
equivalent to GR in a maximal slicing. We show that the tertiary constraint appearing in the spatially closed case has
the general solution $\pi\! =\! a\sqrt{g}$, where $\pi$ is the trace of the momentum conjugate to the three-metric $g_{ij}$, $g$ denotes 
the determinant of the metric and $a(t)$ 
is a function of time. The special choice $a\! =\! 0$ reproduces a gauge-fixed version of GR, but the equivalence proof does not
generalize in an easy way to $a\not= 0$, although we provide some plausibility arguments that it should go through there too. In the
following Sec.\ \ref{accelerate},
attempting to interpret the result of Giulini and Kiefer in terms of the non-projectable $\lambda$-$R$ 
model leads to inconsistencies. We therefore turn in Sec.\ \ref{projectable} to a discussion of the projectable $\lambda$-$R$ model, 
and argue that it exhibits violations of GR as long as $\lambda\neq 1$, in agreement with the argument
put forward in reference \cite{kief}. Our conclusions are contained in Sec.\ \ref{conclusion}.

\subsection{Some elements of Ho\v{r}ava-Lifshitz gravity}
\label{hl_recap}

Ho\v{r}ava-Lifshitz gravity has the ambition to yield a theory of gravity which at high energies remains 
finite and well-defined, while reproducing GR in its classical regime. 
As mentioned in the introduction, HLG realizes a scenario where the renormalization group has a fixed point in the UV, 
at which time (``$t$") and space distances (``$x$") behave asymmetrically under scale transformations.  
More precisely, solutions of the theory at the Planck scale should be compatible with the scaling relations
\begin{equation}
t\rightarrow b^zt ,\qquad x^i\rightarrow bx^i,
\end{equation}
where $b$ is a scaling parameter, $z$ the critical exponent characterizing the fixed point and 
the spatial index takes the values $i=1,2,3$ here and in what follows.
Specific choices of $z$ characterize different models. To obtain a pure gravity theory in $d$ spatial dimensions, 
with up to second-order time derivatives, which is invariant under foliation-preserving diffeomorphisms and 
power-counting renormalizable, one needs $z\! \geq\! d$ (see \cite{hor} for details).

The spacetime manifold in this setting naturally is of the form of a product $\mathds{R}\times\Sigma$, 
where $\mathds{R}$ represents the time direction and $\Sigma$ is a three-dimensional spatial manifold. 
Let $t$ be the time defining the foliation and $x^i$ some coordinates on the spatial hypersurfaces labelled
by $t$. In the presence of a foliation it is convenient to work in a (3+1)-formulation and use the ADM decomposition
of the Lorentzian four-metric in terms of the three-metric $g_{ij}$ on spatial hypersurfaces of constant $t$, 
the lapse function $N$ and the shift vector $N^i$, see \cite{adm} for details. 
The generators of foliation-preserving diffeomorphisms are given by
\begin{equation}
\label{repar}
\delta t=f(t),\qquad \delta x^i=\zeta^i(x,t),
\end{equation}
and act on the ADM field variables according to
\begin{align}
&\delta g_{ij}=\zeta^k\partial_kg_{ij}+f\dot{g}_{ij}+\left(\partial_i\zeta^k\right)g_{jk}+\left(\partial_j\zeta^k\right)g_{ij}\;,\lp
&\delta N_i=\left(\partial_i\zeta^j\right)N_j+\zeta^j\partial_jN_i+\dot{\zeta}^jg_{ij}+\dot{f}N_i+f\dot{N}_i\;,\label{adm} \\
&\delta N=\zeta^j\partial_jN+\dot{f}N+f\dot{N}.\nonumber
\end{align}
Since the infinitesimal generator $f(t)$ of time reparametrizations in \eqref{repar} depends only on $t$ and not on $x^i$,
one possible choice is to let the associated lapse field $N$ also depend on time only. This is different from GR, where the
lapse is a general (positive and nowhere vanishing) function on spacetime.
This ambiguity gives rise to two different versions of Ho\v{r}ava-Lifshitz gravity, {\it projectable} HLG, where $N\! =\! N(t)$, 
and {\it non-projectable} HLG with $N\! =\! N(x,t)$. If one adopts an effective field theory perspective, the latter is more 
complicated to write down since one needs to include in the Lagrangian terms depending on the field
$a_i:= \frac{\partial_i N}{N}$, which turns out to transform like a vector under foliation-preserving diffeomorphisms.

In the present work, instead of considering all possible higher-order terms in spatial derivatives (in $d+1$ spacetime 
dimensions, derivatives up to order $2d$ of the metric are allowed in the potential term), 
we confine ourselves to the HLG-generalization of the terms present in the usual Einstein-Hilbert action 
with a cosmological constant $\Lambda$, namely,
\begin{equation}
S=\int dt\int d^3x\,\sqrt{g}\,N\left(K_{ij}\mathcal{G}^{ijkl}K_{kl}+R-2\Lambda\right),\label{agr}
\end{equation}
where $R$ is the three-dimensional Ricci scalar and 
\begin{equation}
\label{extr}
K_{ij}=\frac{1}{2N}\left(\dot{g}_{ij}-\nabla_iN_j-\nabla_jN_i\right)
\end{equation}
is the extrinsic curvature tensor of the spatial hypermanifolds, with trace $K:=g^{ij}K_{ij}$. 
In writing the action as \eqref{agr}, we have set the overall factor $1/(16\pi G_N)$ depending on Newton's
constant $G_N$ to 1, since it will not play an important role in our classical analysis.
The covariant derivative $\nabla$ in \eqref{extr} is with respect to the three-metric $g_{ij}$, and $\mathcal{G}^{ijkl}$ 
is the Wheeler-DeWitt metric on ``superspace" (the space of all Riemannian three-metrics on $\Sigma$),
\begin{equation}
\mathcal{G}^{ijkl}=\frac{1}{2}\left(g^{ik}g^{jl}+g^{il}g^{jk}\right)-g^{ij}g^{kl}.\label{cor1}
\end{equation}
As mentioned earlier, when reducing the full, four-dimensional diffeomorphism invariance to an invariance under
foliation-preserving diffeomorphisms, the linear combination 
$K_{ij}K^{ij}-K^2$ loses its distinguished character, since both $K_{ij}K^{ij}$ and $K^2$ are now
separately invariant. Consequently, the counterpart in HLG
of the Einstein-Hilbert action \eqref{agr} becomes
\begin{align}
S^\lambda =&\int dt\int d^3x \,\sqrt{g}\,N\left(K_{ij}\mathcal{G}^{ijkl}_\lambda K_{kl}+R-2\Lambda\right)\nonumber \\
=&\int dt\int d^3x\,\sqrt{g}\,N\left(K_{ij}K^{ij}-\lambda K^2+R-2\Lambda\right),\label{ahlg}
\end{align}
where $\lambda$ is a new dimensionless coupling and $\mathcal{G}^{ijkl}_\lambda $ is the generalized 
Wheeler-DeWitt metric
\begin{equation}
\label{wdw}
\mathcal{G}^{ijkl}_\lambda =\frac{1}{2}\left(g^{ik}g^{jl}+g^{il}g^{jk}\right)-\lambda g^{ij}g^{kl},
\end{equation}
which for $\lambda\! =\! 1$ reduces to the standard Wheeler-DeWitt metric \eqref{cor1}. 
In what follows, we will exclude the case $\lambda\! =\!1/3$, for which $\mathcal{G}^{ijkl}_\lambda$ becomes degenerate.
Under this assumption, the inverse of the generalized Wheeler-DeWitt metric exists and is given by
\begin{equation}
\label{wdwinverse}
\mathcal{G}_{ijkl}^\lambda=\frac{1}{2}\left(g_{ik}g_{jl}+g_{il}g_{jk}\right)-\frac{\lambda}{3\lambda-1}g_{ij}g_{kl}.
\end{equation}
We will refer to any theory with action \eqref{ahlg} as a ``$\lambda$-$R$ model", a name coined 
by Bellor\'in and Restuccia in \cite{ven}. An assertion often made in this context is that in order for 
Ho\v{r}ava-Lifshitz gravity to reproduce general relativity in a low-energy limit, the parameter $\lambda$ 
must flow towards its `relativistic' value $\lambda\! =\! 1$, because only then the familiar-looking form of the
action is recovered. 

A different argument for the unphysical nature of $\lambda$-$R$ models for general $\lambda$ was given long
before the advent of HLG by Giulini and Kiefer \cite{kief}. They took the canonical Dirac algebra
of GR's diffeomorphism constraints as their starting point and introduced a $\lambda$-dependent Hamiltonian constraint by
constructing its kinetic part -- the part quadratic in the field momenta $\pi^{ij}$ -- with the help of the
generalized Wheeler-DeWitt metric \eqref{wdw} instead of \eqref{cor1}. By computing particular cosmological observables, and 
finding them to be $\lambda$-dependent, they derived observational constraints on the range of allowed $\lambda$-values,
thereby demonstrating that at least for generic $\lambda$ these models cannot be equivalent to general relativity. 

Note that not all versions of HLG yield \eqref{ahlg} as their lowest-order action; the so-called healthy extensions 
of the non-projectable theory include terms depending on $a_i\equiv\partial_i\log N$ already at this stage \cite{lol}. 
Omitting these terms, it was later argued in \cite{ven} that the resulting non-projectable theory for generic 
values of $\lambda$ reproduces GR. This is at first sight surprising, since the functional form of the action is then {\it not} 
that of the Einstein-Hilbert action, and it also
appears to be in contradiction with the results by Giulini and Kiefer just described.

\section{The non-projectable theory}
\label{np_theory}

The key questions we will address in the remainder of this paper are whether the non-projectable classical 
$\lambda$-$R$ model given by the action \eqref{ahlg} with $N\! =\! N(x,t)$ is equivalent to general relativity for
a compact, three-dimensional manifold $\Sigma$ without boundaries, and whether and how this can be related to
the results derived in \cite{kief}. Our treatment of the non-projectable $\lambda$-$R$ model will follow the
standard Dirac analysis of constrained Hamiltonian systems, as for example described in \cite{Dir,Sund,HT}.

In a nutshell, we will define generalized momenta, read off the resulting primary constraints, and express the Hamiltonian 
functional as a function of the canonical field variables. 
We then impose the consistency condition that the primary constraints be preserved in time, using the symplectic 
(Poisson bracket) structure to compute their time evolution. 
When a time derivative is not weakly equal to zero\footnote{Two phase space functions are said to be weakly equal
when they agree on the constraint surface, the subspace of phase space where all constraints are satisfied.}
we either obtain a secondary constraint 
when the resulting equation only affects the canonical field variables, 
or we can determine a Lagrange multiplier associated with a primary constraint. 
This process is repeated until no new constraints are generated. From that moment on we work in 
the subspace of the phase space defined by the joint vanishing of all constraints $\{ \phi_i, i=1,2, ...,n \}$, the so-called constraint 
surface\footnote{Sometimes a different name is given to the subspace generated at each step. For simplicity, we 
will refer to all of them as ``the constraint surface", implying the space defined at the last step of the algorithm.}. 
We can then classify the constraints into first and second class by
computing the $n\!\times\! n$-matrix $M_{ij}\! =\!\poiss{\phi_i,\phi_j}$ of Poisson brackets between them. 
The rank of $M$ is equal to the number $\mathcal{C}_2$ of second-class constraints,
while the number $\mathcal{C}_1$ of first-class constraints is given by $n-\mathcal{C}_2$. 
This enables us to compute the number $\mathcal{N}$ of local physical degrees of freedom according to
\begin{equation}
\mathcal{N}=\frac{1}{2}\left(\mathcal{P}-2\mathcal{C}_1-\mathcal{C}_2\right),
\end{equation}
where $\mathcal{P}$ is the number of field variables parametrizing the unconstrained phase space.

As already mentioned in the introduction, the time evolution of the Hamiltonian constraint in the non-projectable 
$\lambda$-$R$ model gives rise to a tertiary 
constraint. In the asymptotically flat setting, imposing it was tantamount to the maximal slicing gauge $\pi\! =\! 0$. This makes
the $\lambda$-$R$ model equivalent to general relativity with the same gauge choice, since setting $\pi$ equal to zero
makes all $\lambda$-dependent terms drop out. The result for closed slices we will obtain below is different and
amounts to a tertiary constraint of the form $\pi\! =\! a(t)\sqrt{g}$, with $a(t)$ a function of time. 
This condition is by no means new in the context of general relativity, and known there as ``constant mean curvature gauge". 
It has been studied by York \cite{Y1,Y2} and also lies at the heart of the so-called shape dynamics programme, 
see \cite{SD} and references therein. The observation that surfaces of constant mean curvature appear naturally 
as preferred frames in
low-energy Ho\v{r}ava-Lifshitz gravity has been made earlier in \cite{afsh}. Also \cite{donjac} mentions the
appearance of the constant mean curvature (CMC) gauge in a special case of the Hamiltonian treatment of 
extended HLG. Neither of these references include the CMC gauge condition in a further Dirac constraint analysis,
as we do here.

To make the treatment of the non-projectable model with closed boundary conditions as transparent as possible,
we begin by performing the constraint analysis for general relativity in the presence of the condition $\pi\! =\! a(t)\sqrt{g}$,
before turning to the analogous computation in Ho\v{r}ava-Lifshitz gravity. A good modern review of the Hamiltonian 
formulation of GR in constant mean curvature (CMC) gauge is \cite{num}.

\subsection{General relativity}
Let us first recall the canonical constraint structure of general relativity. Taking as our starting point
the ADM form \eqref{agr} of the Einstein-Hilbert action
we define canonically conjugate, generalized momenta by
\begin{align}
&\pi^{ij}\equiv\frac{\delta S}{\delta\dot{g}_{ij}}=\sqrt{g}\,\mathcal{G}^{ijkl}K_{kl},\label{mom}\\
&\phi_i\equiv\frac{\delta S}{\delta\dot{N}^i}=0,\\
&\phi\equiv\frac{\delta S}{\delta\dot{N}}=0.
\end{align}
We therefore have a total of 20 phase-space variables (or 10 canonical pairs) at each spacetime 
point\footnote{For sake of brevity, we will omit ``at each point'' from now on.}: 12 from $g_{ij}$ and $\pi^{ij}$, 
both of them symmetric three-tensors, 6 from the shift vector $N^i$ and its conjugate momenta $\phi_i$, and 2 
from the lapse function $N$ and its momentum $\phi$. 
The vanishing of $\phi$ and $\phi_i$ defines the four primary constraints of the theory, 
which contribute to the total Hamiltonian
\begin{equation}
\label{ham}H_{tot}=\int d^3x\, \left\{N\left(\frac{\mathcal{G}_{ijkl}}{\sqrt{g}}\pi^{kl}\pi^{ij}-\sqrt{g}\left(R-2\Lambda\right)\right)
-2N^ig_{ik}\nabla_j\pi^{kj}+\alpha^i\phi_i+\alpha\phi\right\}
\end{equation}
with Lagrange multipliers $\alpha$ and $\alpha^i$. The explicit form of the inverse Wheeler-DeWitt metric 
$\mathcal{G}_{ijkl}$ can be obtained from \eqref{wdwinverse} for the special case $\lambda\! =\! 1$.
Demanding that the constraints $\phi\! =\! 0$ and $\phi_i\! =\! 0$ continue to hold under time evolution
implies four secondary constraints,
\begin{align}
&{\cal{H}}:=\poiss{\phi,H_{tot}}=\frac{\mathcal{G}_{ijkl}}{\sqrt{g}}\pi^{kl}\pi^{ij}-\sqrt{g}\left(R-2\Lambda\right)\approx 0,\label{hm}\\
&{\cal H}_i :=\poiss{\phi_i,H_{tot}}=-2g_{ik}\nabla_j\pi^{jk}\approx 0,
\end{align}
where ${\cal H}\!\approx\! 0$ is called the ``Hamiltonian constraint"
and ${\cal H}_i\!\approx\! 0$ are the three ``momentum constraints". Their time preservation does not yield any further constraints 
since both $\dot{{\cal H}}$ and $\dot{{\cal H}_i}$ vanish on the constraint surface. This can be seen from their Poisson bracket
relations
\begin{align}
&\poiss{\int d^3x\, N_1^i{\cal H}_i,\int d^3x'\,N_2^j{\cal H}_j}=\int d^3x\,{\cal H}_i \left(N_1^j\partial_jN^i_2-N_2^j\partial_jN^i_1\right) ,
\label{hihj}\\
&\poiss{\int d^3x\, N_1^i{\cal H}_i,\int d^3x'\,N{\cal H}}=\int d^3x\,{\cal H} N^i\nabla_iN, \label{dir}\\
&\poiss{\int d^3x\, N_1{\cal H},\int d^3x'\,N_2{\cal H}}=\int d^3x\,{\cal H}_ig^{ij} \left(N_1\partial_jN_2-N_2\partial_jN_1\right),
\label{hh}
\end{align}
forming the so-called Dirac algebra of constraints, which has the usual geometric interpretation as a ``projected" version of
the (Lie) algebra of the generators of the four-dimensional diffeomorphism group of the covariant theory. 
Since all eight constraints are first class, the number $\mathcal{N}$ of physical degrees of freedom is given by
\begin{equation}
\mathcal{N}=\frac{1}{2}\left(\mathcal{P}-2\mathcal{C}_1-\mathcal{C}_2\right)=\frac{1}{2}\left(20-16\right)=2,
\end{equation}
which is the usual statement that the gravitational field contains just two local {\it physical} degrees of freedom, 
with the remaining ones being redundant or ``gauge".

\subsection{General relativity in constant mean curvature gauge}
\label{cmcgauge}

Having determined the constraint structure and the equations of motion,
one can proceed by gauge-fixing some of the redundant quantities. 
To pave the ground for the discussion of HLG in the next subsection (where it will appear as a
solution to the tertiary constraint), we will impose the condition 
\begin{equation}
\label{pigauge}
\omega:=\pi-a(t) \sqrt{g} =0,
\end{equation} 
with a (possibly time-dependent) constant $a$. The choice of identifying $a$ with time
is usually referred to as ``York time" \cite{Y1}. 
To make sure the chosen gauge \eqref{pigauge} is preserved in time, the
total time derivative of $\omega$ must vanish also, that is,
\begin{equation}
\frac{d\omega}{dt}= \frac{\partial\omega}{\partial t}+\{ \omega, H_t\}=-\dot{a} \sqrt{g} +
\poiss{\pi-a\sqrt{g},H_{tot}} \approx 0.
\end{equation} 
To simplify calculations, we will from now on assume that $a$ is not time-dependent.
Discarding terms which vanish on the constraint surface, we then obtain the condition
\begin{equation}
\label{mcond}
\mathcal{M}:= \sqrt{g}\left(R- 3 \Lambda +\frac{a^2}{4}-\nabla^2\right)N\approx 0,
\end{equation}
which itself has to be preserved in time. For given $a$, \eqref{mcond} is an elliptic equation for the lapse
function $N$. At this point, we will take advantage of the freedom to
redefine the vector ${\cal H}_i$ of momentum constraints by adding 
a linear combination of constraints to it. More precisely, we will work with momentum constraints of the form
\begin{equation}
{\cal \tilde H}_i=-2g_{ij}\nabla_k\pi^{jk}+ (\nabla_i N)\, \phi ,\label{modmom}
\end{equation}
which alternatively can be viewed as redefining the Lagrange multiplier of
the constraint $\phi$ according to $\alpha \mapsto \alpha + N^i\nabla_i N$.
The motivation for using the constraints ${\cal \tilde H}_i$ is that the original expressions ${\cal H}_i$ only generate 
spatial diffeomorphisms 
of the metric and its momentum. Since the new constraint $\mathcal{M}$ is a functional of the lapse, 
it is necessary to adapt the form of the momentum constraints to make its invariance under spatial diffeomorphisms explicit. 
It is also straightforward to show that the relations (\ref{hihj})-(\ref{hh}) of the Dirac algebra remain unchanged, with
${\cal H}_i$ replaced by ${\cal \tilde H}_i$ everywhere.
Imposing $\dot{\mathcal{M}}\!\approx\! 0$ leads after a lengthy calculation to a differential equation for $\alpha$, namely,
\begin{align}
\label{alpha}
2N\nabla_i\nabla_j N\left(2\pi^{ij}-a\sqrt{g}g^{ij}\right)+(\nabla_iN)(\nabla_jN)\left(2\pi^{ij}-a\sqrt{g}g^{ij}\right)\nonumber\\
+N^2\left(a\sqrt{g}R-2\pi^{ij}R_{ij}\right)+\sqrt{g}\left(R- 3 \Lambda +\frac{a^2}{4}-\nabla^2\right)\alpha\approx 0.
\end{align}
Having imposed all conditions to make sure that the gauge choice is consistent, we can now write down the 
equations of motion for the fields. Without imposing the gauge \eqref{pigauge} we obtain
\begin{align}
\dot{g}_{ij}=&\nabla_i N^kg_{kj}+\nabla_j N^kg_{ik}+\frac{2N}{\sqrt{g}}\mathcal{G}_{ijkl}\pi^{kl},\lp
\dot{\pi}^{ij}=&-N\left(\frac{2g_{kl}}{\sqrt{g}}\left(\pi^{ik}\pi^{jl}-\frac{1}{2}\pi^{kl}\pi^{ij}\right)-\sqrt{g}\left(g^{ij}
\left(R-2\Lambda\right)-R^{ij}\right)\right)\lp
&\qquad + \sqrt{g}\left(g^{ik}g^{jl}-g^{ij}g^{kl}\right)\nabla_k\nabla_l N+\nabla_a\left(N^a\pi^{ij}\right)
-\pi^{a i }\nabla_aN^{ j } -\pi^{a j }\nabla_aN^{ i }       ,\lp[0.5em]
\dot{N}=&\alpha+N^i\nabla_iN,\quad
\dot{\phi}={\cal H}+\nabla_i(N^i\phi)\approx 0,\quad
\dot{N^i}=\alpha^i,\quad
\dot{\phi_i}={\cal \tilde{H}}_i\approx 0.
\end{align}
Substituting $\pi=a\sqrt{g}$, these equations simplify to
\begin{align}
\dot{g}_{ij}=&\nabla_i N^kg_{kj}+\nabla_j N^kg_{ik}+\frac{2N}{\sqrt{g}}\pi_{ij}-aNg_{ij},\lp
\dot{\pi}^{ij}=&-N\left(\frac{2g_{kl}}{\sqrt{g}}\pi^{ik}\pi^{jl}-a\pi^{ij}+\sqrt{g}\left(R^{ij}-g^{ij}\Lambda'\right)\right)+\sqrt{g}g^{ik}g^{jl}
\nabla_k\nabla_l N\lp
&\qquad\qquad +\nabla_a\left(N^a\pi^{ij}\right)  -\pi^{a i }\nabla_aN^{ j } -\pi^{a j }\nabla_aN^{ i }, \lp[0.5em]
\dot{N}=&\alpha+N^i\nabla_iN,\quad 
\dot{\phi}={\cal H}+\nabla_i(N^i\phi)\approx 0,\quad 
\dot{N^i}=\alpha^i,\quad 
\dot{\phi_i}={\cal \tilde{H}}_i\approx 0.
\end{align}
The reason for writing these equations explicitly is to provide a reference point for the corresponding computation
in Ho\v{r}ava-Lifshitz gravity below, where we will see that it is generally not possible to remove their $\lambda$-dependence 
when $a\neq 0$.

\subsection{Ho\v{r}ava-Lifshitz gravity}
\label{hlgrav}

Keeping in mind the results just derived for general relativity, we will now analyze the non-projectable $\lambda$-$R$ model,
whose Ho\v{r}ava-Lifshitz action $S^\lambda$ was given earlier in equation \eqref{ahlg}.
Its Legendre transformation proceeds exactly as before, with $\mathcal{G}^\lambda_{ijkl}$ replacing 
$\mathcal{G}_{ijkl}$ in relations \eqref{mom} and \eqref{ham}. The total Hamiltonian is given by
\begin{align}
H_{tot}^\lambda =&\int d^3x\, \left\{N{\cal H}^\lambda +N^i{\cal \tilde{H}}_i+\alpha^i\phi_i+\alpha\phi\right\},
\end{align}
where the only new quantity is the HLG Hamiltonian constraint ${\cal H}^\lambda$ given by
\begin{equation}
{\cal H}^\lambda := \frac{1}{\sqrt{g} }\, \mathcal{G}_{ijkl}^\lambda\pi^{ij}\pi^{kl}-\sqrt{g}\left(R-2\Lambda\right)\approx 0.
\end{equation}
Since we have not made any changes to the spatial diffeomorphism part of the theory, the constraints ${\cal \tilde{H}}_i$ remain 
unchanged and so does their Poisson bracket algebra \eqref{hihj}. Although the Hamiltonian constraint ${\cal H}$ is replaced
by ${\cal H}^\lambda$, 
it is still invariant under spatial diffeomorphisms and depends on the same fields, which implies that its 
Poisson brackets \eqref{dir} with the momentum constraints are unchanged too.

The only difference with the standard Dirac algebra arises in the $\poiss{{\cal H}^\lambda,{\cal H}^\lambda}$ part of the 
Poisson brackets.  
Omitting the (weakly vanishing) part of this bracket which can be read off from the right-hand side of the
corresponding GR relation \eqref{hh}, one finds
\begin{equation}
\poiss{\int d^3x\, N_1{\cal H}^\lambda,\int d^3x'\,N_2{\cal H}^\lambda}\approx 2\,\frac{1-\lambda}{3\lambda-1}\int d^3z\,\pi\left(N_1\nabla^2N_2-N_2\nabla^2N_1\right), \label{nvbra}
\end{equation}
which does not immediately vanish on the constraint surface unless $\lambda\!=\!1$. 
Demanding $\dot{{\cal H}^\lambda}$ to be weakly equal to zero for general $\lambda$ will therefore generate a tertiary constraint.
Reading off the time evolution of ${\cal H}^\lambda$ from \eqref{nvbra} we obtain
\begin{align}
\poiss{\int d^3x \,N_1{\cal H}^\lambda,\int d^3x'\,N_2{\cal H}^\lambda}&\approx \,\, 2\, 
\frac{1-\lambda}{3\lambda-1}\int d^3z \,N_1\left(\pi\nabla^2N_2-\nabla^2\left(\pi N_2\right)\right),\lp
\Rightarrow\dot{{\cal H}^\lambda}=-2\, \frac{1-\lambda}{3\lambda-1}&\left(N\nabla^2\pi+2g^{ij}
(\nabla_i\pi)(\nabla_jN)\right)\approx 0. \label{tercons}
\end{align}
Note that viewing \eqref{tercons} as a condition for fixing the lapse $N$ as a Lagrange multiplier is not an option here, because
the only possible solution for closed spatial slices would be $N\! =\! 0$, 
which we reject on account of implying a degenerate four-metric. 
To understand better the implications of the new tertiary constraint, let us multiply \eqref{tercons} by $N$ (which by assumption is
non-vanishing), yielding
\begin{equation}
N^2\nabla^2\pi+2g^{ij}N(\nabla_i\pi)(\nabla_jN)=g^{ij}\nabla_i\left(N^2\nabla_j\pi\right)\approx 0.\label{tersol}
\end{equation}
Solutions to \eqref{tersol} can be divided into those with an identically vanishing momentum trace, $\pi\! =\! 0$, 
and those for which $\pi\!\neq\! 0$. Setting $\pi\! =\! 0$ clearly is a solution to the constraint and does not impose any
further restrictions. For nonvanishing $\pi$, we can without loss of generality multiply the equation 
by\footnote{Including the inverse square root of the determinant is necessary to
obtain a quantity of the correct density weight to be integrated.} $\frac{\pi}{\sqrt{g}}$ and integrate it over $\Sigma$, resulting in
\begin{equation}
\int d^3x \,\frac{\pi}{\sqrt{g}}\, g^{ij}\nabla_i\left(N^2\nabla_j\pi\right)=-\int d^3x\,\frac{N^2}
{\sqrt{g}}\, g^{ij}\left(\nabla_i\pi\right)\left(\nabla_j\pi\right)\approx 0.
\end{equation}
Since neither $N$ nor $\sqrt{g}$ are allowed to vanish, we conclude that $\nabla_i\pi$ has to be zero. 
However, this expression can be written as
\begin{equation}
\nabla_i\pi=\sqrt{g}\,\partial_i\left(\frac{\pi}{\sqrt{g}}\right)\approx 0\Rightarrow \pi\approx a\sqrt{g},\label{cmc}
\end{equation}
with $a$ a (possibly time-dependent) constant, proving our earlier assertion of the appearance of the tertiary
constraint \eqref{pigauge} in the context of Ho\v{r}ava-Lifshitz gravity. 
Note that the same condition can be derived in the case of non-compact spatial boundary
conditions, but there $a\! =\! 0$ is forced upon us by the requirement of asymptotic flatness, the special case already
shown to be equivalent to Einstein's theory \cite{ven}.\footnote{This follows from the fall-off conditions on the fields 
implied by the presence of the background flat metric at spatial infinity; 
for $r\rightarrow\infty$ one must have \cite{num}
\begin{equation}
g_{ij}\rightarrow \delta_{ij}+\mathcal{O}(r^{-1}),\qquad \pi^{ij}\rightarrow \mathcal{O}(r^{-2}),\qquad
N\rightarrow 1+\mathcal{O}(r^{-1}),\qquad N^i\rightarrow \mathcal{O}(r^{-1}).
\end{equation}
From these relations, it follows that $\pi\rightarrow\mathcal{O}(r^{-2})$, excluding any choice $a\neq 0$.} 

Repeating the steps of the previous section, we now compute the analogue of 
$\mathcal{M}\approx 0$, which turns out to be
\begin{equation}
\mathcal{M}^\lambda:=
\sqrt{g}\left(R-  3\Lambda+\frac{a^2}{2\left(3\lambda-1\right)} -\nabla^2\right)N
\approx 0.\label{compmcons}
\end{equation}
Demanding that $\mathcal{M}^\lambda$ be preserved in time, and again using the redefined 
momentum constraints ${\cal \tilde H}_i$ of \eqref{modmom}, another lengthy calculation yields 
\begin{align}\label{hlpha}
&2N\nabla_i\nabla_j N\left(2\pi^{ij}-c_na\sqrt{g}g^{ij}\right)+(\nabla_iN)(\nabla_jN)\left(2\pi^{ij}-c_na\sqrt{g}g^{ij}\right)\lp
&\quad +N^2\left(c_r a\sqrt{g}R-2\pi^{ij}R_{ij}\right)+\sqrt{g}\left(R-   3\Lambda+\frac{a^2}{2\left(3\lambda-1\right)}  
 -\nabla^2\right)\alpha\approx 0
\end{align}
as the analogue of condition \eqref{alpha},
where $c_n$ and $c_r$ are $\lambda$-dependent constants given by
\begin{equation}
c_n=\frac{2\lambda-1}{3\lambda-1},\qquad
c_r=\frac{2\lambda}{3\lambda-1}.
\end{equation}
As a cross-check, note that setting $\lambda\! =\! 1$ in \eqref{hlpha} gives back the GR result \eqref{alpha}. 

Let us summarize what we have learned about the constraint structure of the theory. 
There are six first-class constraints, ${\cal H}_i\!\approx\! 0$ and $\phi_i\!\approx\! 0$, 
and four second-class constraints, ${\cal H}^\lambda\!\approx\! 0$, $\mathcal{M}^\lambda\!\approx\! 0$, 
$\phi\!\approx\! 0$ and $\omega\!\approx\! 0$, bringing the total number of physical degrees of freedom to two, just like
in general relativity. There are other close parallels with GR in constant mean curvature gauge: the field content is
the same, the spatial diffeomorphisms and their associated three primary and three secondary first-class constraints
coincide, and there is a one-to-one correspondence between the conditions imposed on the fields. 
However, due to the non-vanishing of the terms proportional to $\pi$ in the Hamiltonian, the equations of motion 
are explicitly $\lambda$-dependent and {\it not} obviously equivalent to those of gravity. Explicitly, they are
\begin{align}
\dot{g}_{ij}=&\nabla_i N^kg_{kj}+\nabla_j N^kg_{ik}+\frac{2N}{\sqrt{g}}\pi_{ij}-2a\frac{\lambda}{3\lambda-1}Ng_{ij},\lp
\dot{\pi}^{ij}=&-N \bigg(   \frac{2g_{kl}}{\sqrt{g}}\pi^{ik}\pi^{jl}-2 a \frac{\lambda}{3\lambda-1}\pi^{ij}-
\frac{\pi_{kl}\pi^{kl}}{2\sqrt{g}}g^{ij} \lp
&\qquad\qquad +\sqrt{g}\left( R^{ij}-g^{ij} 
\left( 2\Lambda-\frac{\lambda a^2}{3\lambda-1}-\frac{R}{2}\right)\right)\bigg) \lp[0.5em]
&\qquad\qquad+\sqrt{g}g^{ik}g^{jl}\nabla_k\nabla_l N+\nabla_a\left(N^a\pi^{ij}\right) 
-\pi^{a i }\nabla_aN^{ j } -\pi^{a j }\nabla_aN^{ i }, \lp[0.7em]
=&-N\left(\frac{2g_{kl}}{\sqrt{g}}\pi^{ik}\pi^{jl}-2a\frac{\lambda}{3\lambda-1}\pi^{ij}
+\sqrt{g}\left(R^{ij}-g^{ij} \left(  \Lambda -\frac{\lambda a^2}{2(3\lambda-1)}\right)   \right)\right)\lp[0.5em]
&\qquad\qquad+\sqrt{g}g^{ik}g^{jl}\nabla_k\nabla_l N+\nabla_a\left(N^a\pi^{ij}\right)
-\pi^{a i }\nabla_aN^{ j } -\pi^{a j }\nabla_aN^{ i }, \lp[0.7em]
\dot{N}=&\alpha+N^i\nabla_iN,\;\;\;\;
\dot{\phi}={\cal H}+\nabla_i(N^i\phi)\approx 0,\;\;\;\;
\dot{N^i}=\alpha^i,\;\;\;\;
\dot{\phi_i}={\cal \tilde{H}}_i\approx 0.
\end{align}
Part of the $\lambda$-dependence can be absorbed into $a$ by setting $\tilde{a}=2a\frac{\lambda}{3\lambda-1}$
and redefining the cosmological constant by a $\lambda$- (and $a$-)dependent term, but this does not in any obvious way
eliminate the $\lambda$-dependence of $N$ and $\alpha$ inherent in equations \eqref{compmcons} and
\eqref{hlpha}. Of course, to show that physics depends on $\lambda$, we would have to exhibit a 
$\lambda$-dependent {\it observable}, in the spirit of Giulini and Kiefer \cite{kief}. We will look in the next section
at their physicality criterion and find that it cannot be applied in a straightforward way to the case at hand.
On the basis of this observation and the great overall similarity with the case $\pi \! =\! 0$ (which is the {\it generic}
solution for the tertiary constraint of the non-projectable $\lambda$-$R$ model with non-compact slices), 
including the counting of local degrees of freedom, we conjecture that the theory with closed slices is also
equivalent to general relativity, although we have not yet been able to show this explicitly by demonstrating that 
the residual $\lambda$-dependence is pure gauge.

\section{Acceleration of the three-volume and projectability}
\label{accelerate}

Having derived the tertiary constraint for the non-projectable theory for closed spatial slices, 
we would like to see under which conditions, if any, the result of \cite{kief} can be reproduced. 
In \cite{kief}, in search of a physical observable, the acceleration of the spatial volume 
\begin{equation}
V(t)=\int d^3x\sqrt{g}\label{vol}
\end{equation}
of the universe was calculated, where ``time" $t$ refers to proper time in the so-called 
proper-time (or canonical) gauge $N\! =\! 1$, $N^i\! = \! 0$. 
The result found for pure gravity (without matter) according to \cite{kief} is
\begin{equation}
\ddot{V}(t)=-\frac{2}{3\lambda-1}\int d^3x\,\sqrt{g}\left(R-3\Lambda\right),\label{kgvol}
\end{equation}
and therefore explicitly $\lambda$-dependent. For $\lambda\! = \! 1$, it reduces to
\begin{equation}
\ddot{V}(t)=-\int d^3x\sqrt{g}\left(R-3\Lambda\right),
\label{accgr}
\end{equation}
matching the classical computation for general relativity \cite{bdw}. The formulas for the volume and
its acceleration still refer to a specific set of gauge-fixed coordinates, but it is argued in \cite{kief} that the 
{\it sign} of the acceleration does not and is a bona fide observable in the sense of being defined invariantly.
In general relativity, a positive scalar curvature $R$ contributes negatively to the acceleration \eqref{accgr},
while a positive cosmological constant contributes positively, both being familiar features of standard
cosmology. However, the sign of the prefactor of the integral of $(R-3\Lambda)$ in equation \eqref{kgvol} 
depends explicitly on the value of $\lambda$. Giulini and Kiefer
point out that this has potential cosmological consequences, at the very least implying bounds on the
allowed values of $\lambda$.
As we will show next, a similar conclusion cannot generally be obtained in the $\lambda$-$R$ model 
derived from non-projectable HLG which we studied in the last section.

Recall that Ho\v{r}ava-Lifshitz gravity from the outset works with a preferred time foliation, where each spatial hypersurface 
of constant time is one of the leaves of the foliation. Its Hamiltonian formulation looks very much like
general relativity in terms of ADM variables, but the role of the lapse is different; in GR the foliation is still arbitrary, 
which is reflected in the presence of the full Dirac algebra \eqref{hihj}-\eqref{hh} of constraints 
and the fact that the Lagrange multiplier $N$ can in principle be any (strictly positive) function.  
By contrast, as we have already seen in Sec.\ \ref{hlgrav}, in Ho\v{r}ava-Lifshitz gravity $N$ cannot be chosen freely,
since the full four-dimensional diffeomorphism symmetry is not only non-manifest, but no longer present.

While we cannot generally fix the lapse to $1$ due to the extra constraints on the HLG model, 
nothing prevents us, at least locally, from choosing a vanishing shift vector, $N^i\! =\! 0$. 
Under this assumption, we can compute $\ddot{V}$ in a straightforward manner to obtain
\begin{align}
&\dot{V}=-\frac{1}{3\lambda-1}\int d^3x N\pi=-\frac{a}{3\lambda-1}\int d^3x N\sqrt{g},\lp
\Rightarrow\; &\ddot{V}=\frac{a}{3\lambda-1}\int d^3x\sqrt{g}\left(\frac{a}{3\lambda-1}N^2-\alpha\right),\label{npacc}
\end{align}
where for simplicity we still assume that $a$ is a constant, as in previous sections.
Using \eqref{compmcons} we can rewrite \eqref{npacc} as
\begin{equation}
\ddot{V}=-\frac{2}{3\lambda-1}\int d^3x\sqrt{g}\left(N \left(R-3\Lambda-\nabla^2\right)N+\frac{a}{2}\alpha\right).\label{tnpacc}
\end{equation}
While the pre-factor is the same as in \eqref{kgvol}, one cannot make general statements about the sign of the acceleration,
since both $\alpha$ and $N$ are $\lambda$-dependent, by virtue of the conditions \eqref{compmcons} and \eqref{hlpha}.
If we could consistently set $\alpha\! =\! 0$ and $N\! =\!1$, the arguments of \cite{kief} would carry over to the present case, 
but again this does not appear to be possible, since both \eqref{compmcons} and \eqref{hlpha}
require a nontrivial functional dependence of $\alpha$ and $N$ on the other field variables, at least for the case 
$a\! =\! const.$ considered here.

\subsection{Proper-time gauge in the non-projectable theory}
\label{noprojpt}

We saw in the previous section that preserving the constraint $\omega\equiv\pi\! - a\sqrt{g} \approx\! 0$ in time yields an equation 
for $N$ which in general is incompatible with the gauge choice $N\! =\! 1$ and $N^i\! =\! 0$. 
We will show in this section that imposing proper-time gauge in the non-projectable theory from the start 
makes little difference to the discussion of Sec.\ \ref{hlgrav}; one is again led to a CMC condition, and a contradiction arises.   
Going back to the action \eqref{ahlg}, nothing seems to prevent us from choosing the proper-time gauge
$N\! =\! 1$, $N^i\! =\! 0$, as long as we 
impose by hand the equations we would otherwise have obtained from varying with respect to $N$ and $N^i$.
Since nothing out of the ordinary happens with the shift part of the gauge, let us focus on the lapse. 
It is straightforward to see that under the Legendre transformation, imposing the Euler-Lagrange equation of the lapse 
is equivalent to the modified Hamiltonian constraint ${\cal H}^\lambda\! \approx\! 0$, which then appears as a primary constraint 
of the theory, together with ${\cal H}_i\! \approx\! 0$. The total `proper-time' Hamiltonian is given by
\begin{equation}
H_{tot}^{pt}=\int d^3x\,\left({\cal H}^\lambda+\gamma{\cal H}^\lambda+\gamma^i{\cal H}_i\right),\label{hamcan}
\end{equation}
where only the first instance of ${\cal H}^\lambda$ comes directly from the Legendre transformation, 
both ${\cal H}^\lambda=0$ and ${\cal H}_i=0$ are primary constraints, and $\gamma$ and $\gamma^i$ 
are Lagrange multipliers. 

In order to allow for an interpretation of $H_{tot}^{pt}$ in terms of four-geometry, the Lagrange multiplier $\gamma$
must be such that $\gamma+1$ is strictly positive and therefore can be interpreted as a lapse function. 
Other than this, $\gamma$ should at this stage be freely specifiable. 
From this point on, the analysis of the problem proceeds along the lines of Sec.\ \ref{hlgrav}, 
with $N$ replaced by $\beta\! :=\!\gamma+ 1$ both in the time evolution of ${\cal H}^\lambda$ 
and in the constraint $\mathcal{M}^\lambda\!\approx\! 0$. No new computations are required, 
because $\beta$ is already a Lagrange multiplier and, due to $\mathcal{M}^\lambda\approx 0$, depends on 
$\lambda$. The analogue of equation \eqref{npacc} now reads
\begin{align}
&\dot{V}=-\frac{1}{3\lambda-1}\int d^3x\, \beta\pi=-\frac{a}{3\lambda-1}\int d^3x\, \beta \sqrt{g},\lp
\Rightarrow\; &\ddot{V}=\frac{a}{3\lambda-1}\int d^3x\,\sqrt{g} \, \frac{a}{3\lambda-1}\,\beta^2 =
\int d^3x\,\sqrt{g} \left( \frac{a}{3\lambda-1}\,\beta \right)^2.\label{newnpacc}
\end{align}
Because of the absence of the term proportional to $\alpha$ that was present in the previous expression
\eqref{npacc}, we have been able to rewrite the integrand on the right-hand side of the acceleration as a
square, which means that the acceleration has to be positive and vanishes only when $a$ is identically zero.
However, as can be seen from relation \eqref{accgr}, there are no such restrictions on the sign of the acceleration of
the spatial volume in standard gravity cosmology. It simply means that in non-projectable
Ho\v{r}ava-Lifshitz gravity, no matter what
the value of little lambda, the choice of proper-time gauge, and of $N\! =\! 1$ in particular, is inconsistent.

The main obstacle to rederiving the results of \cite{kief} is the presence of the Hamiltonian constraint ${\cal H}^\lambda$
which does not Poisson-commute with itself on the constraint surface. 
In search of alternative derivations, let us briefly investigate how far we can get when dropping the Hamiltonian
constraint  ${\cal H}^\lambda \approx\! 0$ altogether. This corresponds to a $\lambda$-$R$ model without any
time reparametrization invariance, where only the spatial diffeomorphisms act as gauge transformations. (For the time
being, we will not bother to analyze how this affects the counting of physical degrees of freedom of the model.)
From the point of view of the action, it amounts to setting $N\! =\! 1$ without any further restrictions. 
Performing the Legendre transformation, we obtain as Hamiltonian the expression \eqref{hamcan} with $\alpha\! =\! 0$, that is,
\begin{equation}
H=\int d^3x\,\left({\cal H}^\lambda +\alpha^i{\cal H}_i\right).
\label{hamH}
\end{equation}
We would like to stress that despite using ${\cal H}^\lambda$ as a shorthand for the functional 
$\frac{\pi^{ij}\pi^{kl}}{\sqrt{g}}\mathcal{G}^\lambda_{ijkl}-\sqrt{g}\left(R-2\Lambda\right)$, 
this model has only momentum constraints and {\it no} Hamiltonian constraint.

It is straightforward to check that taking Poisson brackets of the momentum
constraints with the Hamiltonian \eqref{hamH} -- the usual consistency check for constraints -- does
not give rise to any kind of Hamiltonian constraint. The relevant Poisson bracket relation can be read off
relation \eqref{dir} for the special case $N\! =\! 1$. Although the integrand on the right-hand side is still proportional 
to ${\cal H}^\lambda$, it is at the same time seen to be a total derivative. Consequently, the time evolution of
the momentum constraints, computed with the Hamiltonian \eqref{hamH},
vanishes without generating any new constraints. 

We note in passing that if instead of $N\! =\! 1$ we would have a space-independent smearing function $N\! =\! N(t)$
(as will be the case in the projectable theory in the following section), the same conclusion would apply, namely,
\begin{equation}
\poiss{\int d^3x\, N^i{\cal H}_i,\int d^3x'\, N(t){\cal H}^\lambda}=0.
\end{equation}

Having established that no new constraints arise, the time derivative of the three-volume is simply given by
\begin{equation}
\dot{V}=\poiss{\int d^3x\,\sqrt{g},H}=-\frac{1}{3\lambda-1}\int d^3x\,\pi.
\end{equation}
Taking another time derivative we obtain the acceleration
\begin{equation}
\ddot{V}=\frac{1}{3\lambda-1}\int d^3x\,\left(\frac{3}{2}
\left(-\frac{\mathcal{G}_{ijkl}^\lambda}{\sqrt{g}}\pi^{ij}\pi^{kl}+2\sqrt{g}\Lambda\right)-\sqrt{g}\,\frac{R}{2}\right).\label{acce}
\end{equation}
Comparing again to the results of \cite{kief}, because of the absence of a Hamiltonian {\it constraint} in the present case, 
it is not possible to rewrite the integrand of \eqref{acce} to obtain an expression depending on the three-dimensional
Ricci scalar and a cosmological constant, as in relation \eqref{kgvol}. However, as we will see in Sec.\ \ref{projectable} below,
there is a $\lambda$-$R$ model which has a Hamiltonian constraint -- albeit a global one -- and no tertiary constraint, 
and which precisely realizes the Giulini-Kiefer scenario.

\subsection{Projectable $\lambda$-$R$ model}
\label{projectable}

Having exhausted all of the potentially relevant variants of the non-projectable $\lambda$-$R$ model, we now turn to the model
derived from the projectable version of Ho\v{r}ava-Lifshitz gravity. As we have explained in Sec.\ \ref{hl_recap},
the non-projectable theory has a general lapse function $N\! =\! N(x,t)$, whereas the projectable one is characterized
by $N\! =\! N(t)$. In light of this, the action \eqref{ahlg} in the projectable case becomes
\begin{equation}
S^\lambda_{pr}=\int dt\,N\int d^3x\, \sqrt{g}\left(K_{ij}\mathcal{G}^{ijkl}_\lambda K_{kl}+R-2\Lambda\right),
\end{equation}
where we have taken $N$ outside the spatial integral to highlight its independence of spatial coordinates. 
Unlike what happened in the non-projectable case, $S^\lambda_{pr}$ really is the most general 
second-order action in spatial derivatives in this version of the theory.
Its Legendre transformation can be performed in a straightforward manner, taking into account that the primary 
constraint $\phi(t)$ defined by the vanishing of the momentum of the lapse $N(t)$ will also depend on time only.

Keeping our previous notation for the functional form of ${\cal H}^\lambda$ and ${\cal H}_i$ -- without at this stage
making any assumption on their constrained character -- the total Hamiltonian takes the form
\begin{equation}
H_{tot}^{pr}= \alpha\phi +N\int d^3x\,{\cal H}^\lambda  + \int d^3x\left(N^i{\cal H}_i+\alpha^i\phi_i\right).
\end{equation}
To obtain the secondary constraints, we must impose that the primary constraints be preserved in time, leading to
\begin{align}
&\quad\; \poiss{\phi(t),H_{tot}^{pr}}\approx 0\;\; \Rightarrow\;\; \int d^3x\,{\cal H}^\lambda \approx 0,\lp
&\poiss{\phi_i(x,t),H_{tot}^{pr}}\approx 0\;\; \Rightarrow\;\;\;\; {\cal H}_i\approx 0.
\end{align}
Not having made any changes to the action of the spatial diffeomorphisms, we obtain the usual momentum constraints,
but instead of the usual Hamiltonian constraint (one at each point $x$), there is only a single, integrated Hamiltonian
constraint at each fixed time $t$, reflecting the reduced dependence of the lapse. 
It is precisely this feature which will allow us to rederive the results of \cite{kief}.

We must show next that the secondary constraints are preserved in time. 
It turns out that all relevant computations have already been done in earlier sections.
The analogue of the Dirac algebra in the present case has the momentum constraints and their Poisson brackets
\eqref{hihj} unchanged. The counterparts of relations \eqref{dir}, that is, the Poisson brackets of the {\it integrated} 
Hamiltonian constraint 
$\int d^3x\,{\cal H}^\lambda\! \approx\! 0$ with the local momentum constraints vanish identically, as we have
already argued in Sec.\ \ref{noprojpt} above. Lastly, the Poisson brackets of the integrated Hamiltonian
with itself vanish also, since all non-vanishing contributions to the right-hand side of relation \eqref{hh} are associated
with a non-trivial spatial dependence of the lapse functions. We conclude that on the constraint surface
the secondary constraints of the projectable $\lambda$-$R$ model are automatically preserved in time and 
no tertiary constraints arise.
  
Given the previous arguments, we are free to set $N\! =\! 1$ from the beginning, in which case the acceleration of the 
three-volume $V$ reduces to our previous formula \eqref{acce}. The difference here is that a genuine Hamiltonian
constraint is present, albeit an integrated one, $\int d^3x\, {\cal H}^\lambda\! =\! 0$. 
This allows us to perform the simplification we previously could not apply to the right-hand side of equation \eqref{acce}
to obtain
\begin{align}
\ddot{V}= \poiss{\dot{V},H}=&\,\frac{1}{3\lambda-1}\int d^3x\left(\frac{3}{2}\left(-\frac{G_{ijkl}}{\sqrt{g}}\pi^{ij}\pi^{kl}
+2\sqrt{g}\,\Lambda\right)-\frac{1}{2}\sqrt{g}\,R\right)\lp
=&\,\frac{1}{3\lambda-1}\int d^3x\left(-\frac{3}{2}{\cal H}-\frac{3}{2}\sqrt{g}\,R+3\sqrt{g}\,\Lambda
+3\sqrt{g}\,\Lambda-\frac{1}{2}\sqrt{g}\,R\right)\lp
=&-\frac{2}{3\lambda-1}\int d^3x\,\sqrt{g}\left(R-3\Lambda\right),
\end{align}
which is exactly the desired result from \cite{kief}. We may then take over their conclusion that at least for some range of
the parameter $\lambda$, the classical predictions of the projectable $\lambda$-$R$ model are not compatible with standard
cosmology and therefore inequivalent to general relativity. This is potentially interesting in its own right, because it 
provides an additional physical criterion on whether projectable Ho\v{r}ava-Lifshitz gravity can be a viable theory,
about which there is some doubt in view of the fact that compared with
standard gravity it has an additional scalar degree of freedom (see, for example, the reviews \cite{wein,muko} 
and references therein for related criticism).

\section{Summary and conclusions}
\label{conclusion}

Our analysis was motivated by two apparently contradictory claims about the classical equivalence between general
relativity and the so-called $\lambda$-R models associated with Ho\v{r}ava-Lifshitz gravity. These are models 
described by an action of the form
\begin{equation}
S=\int dt \int d^3x\,\sqrt{g}N\left(K_{ij}K^{ij}-\lambda K^2+R-2\Lambda\right),
\end{equation}
which differs from the standard gravitational Einstein-Hilbert action through its dependence on the real parameter
$\lambda$. By performing a Hamiltonian constraint analysis \`a la Dirac, we have shown that there is no 
contradiction after all. The work by Giulini and Kiefer cannot be interpreted consistently in the framework of
the non-projectable version of HLG because the gauge choice made in \cite{kief} is not compatible with the
structure of the constraint algebra in non-projectable Ho\v{r}ava-Lifshitz gravity 
and the consistency requirements following from it. By contrast, the results of Bellor\'in and Restuccia in \cite{ven}  
were obtained for the non-projectable $\lambda$-$R$ model, and crucially relied on the spacetime dependence $N(t,x)$ 
of the lapse function. 

In order to study the behaviour of the total three-volume of the universe introduced in \cite{kief} and make the comparison 
between the two formulations explicit, 
we had to repeat the Dirac analysis of reference \cite{ven} for closed spatial slices. 
We found the same tertiary constraint, but because of the different boundary conditions the class of allowed solutions
was larger and of the form $\pi=a\sqrt{g}$, with $a$ a (possibly time-dependent) constant. 
This raised the question of whether the enlarged solution set still leads to theories equivalent to GR (in constant mean
curvature gauge), in the same way that the unique solution $\pi\! =\! 0$ for the asymptotically flat case
can be shown to be equivalent to GR in a maximal slicing gauge. Because of the involved nature of the
$\lambda$-dependence of the consistency conditions arising in the case of closed slices, we were unable
to show that the $\lambda$-dependence is pure gauge. However, we think that this is plausible, given the similarity
with general relativity in CMC gauge otherwise. An alternative possibility would be that choices $a(t)\! \not= \! 0$ for some
reason are inconsistent, which would again leave us with $\pi\! =\! 0$ as the only solution to the tertiary
constraint. 

Even if the equivalence with general relativity of the non-projectable $\lambda$-$R$ model for closed spatial 
slices cannot be shown to hold beyond $a\! =\! 0$, we would like to re-iterate a point already made in \cite{ven}, namely,
that $\lambda\! \not= \! 1$ does not necessarily indicate a deviation from general relativity. Requiring
$\lambda$ to go to 1 when considering the low-energy limit of non-projectable Ho\v{r}ava-Lifshitz gravity may 
therefore be too restrictive. Since renormalization group computations have been initiated in the 
context of HLG \cite{AF,DB}, it will be interesting to see what flow is realized for $\lambda$ and whether it is 
possible to recover some partially gauge-fixed version of the theory while avoiding the strong-coupling pitfalls that have been 
shown to occur when $\lambda\rightarrow 1$ \cite{Str,Str2,Str3}. 

The question of the role and physical interpretation of little lambda is also important
in the Causal Dynamical Triangulations (CDT) approach to quantum gravity, which can accommodate some of
the anisotropic features of HLG \cite{cdthlg}. 
A main result in CDT quantum gravity is the fact that a minisuperspace version $S^{\mathit{eff}}$
of the action \eqref{ahlg} turns out to govern the dynamics of the three-volume of the universe. Nonperturbative contributions  
to the corresponding coupling $\lambda^{\mathit{eff}}$ are crucial in bringing about a classical limit compatible with 
GR \cite{cdt} (see also \cite{budd,and} for investigations of $\lambda$ in the context of CDT in three spacetime dimensions).

Our investigation has highlighted 
that the role of ``little $\lambda$" is rather subtle, even in the classical theory, and depends on the
precise model one is looking at. To understand the constraint structure of
the theory and why the construction of \cite{kief} is inconsistent with non-projectable HLG, we had to perform 
the Dirac analysis from the beginning. The strategy to look at the acceleration of the three-volume 
in the framework of non-projectable HLG was inconclusive, even when we dropped the time reparametrization invariance 
and associated Hamiltonian constraint altogether. Imposing proper-time gauge from the outset, as in reference
\cite{kief}, turned out to be inconsistent with the CMC condition which is still necessary to close the constraint algebra.
However, we have found that {\it projectable} HLG can
accommodate both the computation and conclusions of \cite{kief}, if one makes the implicit assumption that the lapse
function is only time-dependent and the Hamiltonian constraint is therefore a single, global condition. With this interpretation,
there is no contradiction between the results of \cite{kief} and \cite{ven}. 

Although all of our computations were for pure gravity with a cosmological constant, the conclusions do not change when 
we include an ultra-local matter term in the Hamiltonian constraint, similar to what was done in \cite{kief}. 
Given its ultra-local nature, it would not play any role in the existence of the tertiary constraint and, at least for $a\! =\! 0$, 
the theory would still be equivalent to GR in maximal slicing gauge.

\vspace{.5cm}

\noindent {\bf Acknowledgements.} 
LP acknowledges financial support from Funda\c{c}\~{a}o para a Ci\^{e}ncia e Tecnologia, Portugal through grant no. SFRH//BD/76630/2011 and thanks the Perimeter Institute for hospitality. Both authors thank S. Carlip, D. Giulini and C. Kiefer for 
discussion during various stages of this work.

\end{document}